\documentclass[12pt]{article}
\usepackage[dvips]{graphicx}
\usepackage{pdproc}

  \makeatletter 
  \def\@cite#1{[#1]} 
  \makeatother    
\def\kt{\mbox{\boldmath{$k$}}}

\begin{document}

\renewcommand{\thefootnote}{\alph{footnote}}

\title{
Supersymmetric Contributions to the $B \to \phi K$ Decays\\
in the PQCD Approach~\footnote{Talk presented at {\it the 12th
International Conference on Supersymmetry and Unification of Fundamental
Interactions (SUSY 2004)}, Tsukuba, Japan, June 17-23, 2004.
This talk is based on the work done in
Ref.~\cite{Mishima:2003ta} collaborated with A.~I.~Sanda.} 
}

\author{SATOSHI MISHIMA}

\address{ 
Department of Physics, Tohoku University, Sendai 980-8578, Japan
\\ {\rm E-mail: mishima@tuhep.phys.tohoku.ac.jp}}

\abstract{
We study the effects of supersymmetric contribution on both 
the $B_d\to\phi K^0$ and $B^\pm \to \phi K^\pm$ modes using 
the perturbative QCD approach.  
We estimate the deviation of mixing-induced and direct CP
asymmetries and discuss the strong-phase dependence of them.   
}

\normalsize\baselineskip=15pt

\section{Introduction}
The CP asymmetry of the $B_d \to \phi K^0$ mode may be useful in the
search for new physics beyond the standard model (SM), since it is
induced only at the one-loop level.  In the SM, the mixing-induced CP
asymmetry, denoted by $S_{\phi K^0}$, must be equal to $\sin(2\phi_1)$,
which is measured from the CP asymmetry of charmonium modes, and the
direct CP asymmetry, $A_{\phi K^0}$, vanishes.  Any difference between
$S_{\phi K^0}$ and $\sin(2\phi_1)$ would be a signal for new physics.
Before the summer in 2004 the Belle collaboration had reported an large
anomaly, $S_{\phi K_S} = -0.96\pm 0.50\, {}^{+\, 0.09}_{-\, 0.11}$,
while the BaBar result had been 
$S_{\phi K^0} = 0.47\pm 0.34\, {}^{+\, 0.08}_{-\, 0.06}$ in agreement
with $\sin(2\phi_1)$~\cite{Aubert:2004ii,Abe:2003yt}.  In the summer in
2004, the both collaborations have given the new
results~\cite{Aubert:2004dy,Sakai:ichep2004}: 
$S_{\phi K^0} = 0.50\pm 0.25\, {}^{+\, 0.07}_{-\, 0.04}$ (BaBar), 
$0.06\pm 0.33\pm 0.09$ (Belle), and 
$A_{\phi K^0} = 0.00\pm 0.23\pm 0.05$ (BaBar), 
$0.08\pm 0.22\pm 0.09$ (Belle).  Although the new Belle data of 
$S_{\phi K^0}$ has moved toward close to the SM value, the data of
$S_{\phi K^0}$ seem to be somewhat smaller than $\sin(2\phi_1)$.  It
might be the effect of some new physics on the $b\to s$
penguin.  Supersymmetry (SUSY) is an attractive candidate for new
physics at TeV scale, thus we would like to study the SUSY contribution
in the $B\to \phi K$ modes.

We analyze the SUSY contribution using the mass insertion approximation
(MIA), which is a powerful technique for model-independent analysis of
new physics associated with the minimal supersymmetric standard model
(MSSM).  In this approximation, the squark propagators with the 
$\tilde b\to \tilde s$ transition can be expanded as a series in terms
of $(\delta^d_{AB})_{23}=(m^2_{\tilde d_{AB}})_{23}/m^2_{\tilde q}$,
where $m^2_{\tilde d}$ is the squared down-type-squark mass matrix, 
$m_{\tilde q}$ an averaged squark mass. 
\{$A$, $B$\} indicate \{$L$, $R$\}, which refer to the helicity of
sfermions.

A problem lies in the evaluation of hadronic matrix elements. 
The CP asymmetries, both the mixing-induced and direct ones, depend on
the strong phase which is generated from the final-state interactions.
However, it is difficult to calculate the decay amplitude including the
strong phase. 
To calculate it, there are several approaches, for example, 
perturbative QCD (PQCD)~\cite{Keum:2000ph}, 
QCD factorization (QCDF)~\cite{Beneke:1999br}, and so on.  
PQCD is based on $k_T$ factorization~\cite{Nagashima:2002ia}, on the
other hand, QCDF on collinear factorization.  
Each method is plagued with large theoretical uncertainties.  
In this talk, we use the PQCD approach to estimate
the MSSM contribution in both the $B_d\to\phi K^0$ and 
$B^\pm \to \phi K^\pm$ modes, and discuss the strong-phase dependence of
the results.

\section{PQCD Approach for $B\to\phi K$}
A key ingredient of the PQCD approach is the factorization of decay
amplitudes into a multiplication of long-distant part and short-distant
part. 
A typical decay amplitude for $B\to\phi K$ can be expressed as the
convolution of a hard part $H$, meson wave functions $\Phi_M$'s and a
Wilson coefficient $C$: 
\begin{eqnarray}
{\cal M} 
  &=& 
    \int_0^1 dx_1 dx_2 dx_3 
    \int_0^\infty b_1db_{1}\, b_2db_{2}\, b_3db_{3}\, 
    \Phi_K(x_2,b_{2})\, e^{-S_K(x_2,b_2,t)}\, 
    \Phi_\phi(x_3,b_{3})\, e^{-S_\phi(x_3,b_3,t)}
  \nonumber\\
  & &\hspace{17mm}\times
    C\left(t\right)  H(x_1,x_2,x_3,b_{1},b_{2},b_{3},t)\,
    \Phi_B(x_1,b_{1})\, e^{-S_B(x_1,b_1,t)}
  \,,
\label{eq:ktfac}
\end{eqnarray}
where $x_i$ and $b_i$ are the longitudinal fraction of partons
momenta and the conjugate variable to the transverse components of
them, respectively.  The scale $t$ is of order of 
$\sqrt{\bar\Lambda M_B}$ with $\bar\Lambda=M_B-m_b$.  Here $S_M$
denotes the Sudakov factor.  The Sudakov factor ensures the absence of
the end-point singularities, thus the arbitrary cutoffs used in QCDF are
not necessary in PQCD.  As a result, we can predict not only the
factorizable contributions but also non-factorizable and annihilation
ones, which cannot be calculated in the naive factorization method.  
A large strong phase is induced from absorptive part in the
annihilation diagrams~\cite{Keum:2000ph}.  Although PQCD has large
theoretical uncertainties, most of them are expected to be canceled in
the ratio when we consider the CP asymmetries.  Here we neglect the
errors coming from the PQCD method.

PQCD has been applied to the leading-order amplitudes of the 
$B\to \phi K$ decays~\cite{Chen:2001pr} and to the chromo-magnetic
penguin (CMP) amplitudes of them~\cite{Mishima:2003wm} within the SM. 
SUSY contribution comes through the CMP amplitude, which is ambiguous in 
the naive factorization method because the magnitude of the momentum
transferred $q^2$ by the gluon in the CMP is unknown.  In the PQCD and
QCDF methods, the CMP can be calculated without any assumption for the
value of $q^2$.  The CMP generates a strong phase from its absorptive
part in PQCD~\cite{Mishima:2003wm}, since $q^2$ is written as
$(1-x_2)x_3M_B^2-|\kt_{2T}-\kt_{3T}|^2$, where $\kt_{iT}$'s are
transverse momenta of the partons, and $q^2$ can vanish.  On the other
hand, in QCDF, $q^2$ can be written in terms of the momentum fraction of
partons too, they however neglect the transverse momenta so that
$q^2=(1-x_2)x_3M_B^2$ and $q^2$ never vanishes.  Therefore, there is no
absorptive part in the CMP amplitude and the strong phase is not
generated from it in contrast with the case of PQCD.

\section{MSSM Effects on $B\to \phi K$}
We estimate the gluino contribution to the CP asymmetries for both
$B_d\to\phi K^0$ and $B^\pm \to \phi K^\pm$ in the single mass-insertion
scheme.  The $LR$ insertion may change the CP asymmetries of 
$B\to\phi K$ significantly even when we constrain the MIA parameters
from the branching ratio of $B\to X_s\gamma$. In the following study, we
take a somewhat conservative bound, 
$2.5 \times 10^{-4}< {\rm Br}(B\to X_s\gamma) < 4.1 \times 10^{-4}$, 
and the soft masses to be $500$ GeV.

\begin{figure}[t]
\begin{center}
\includegraphics*{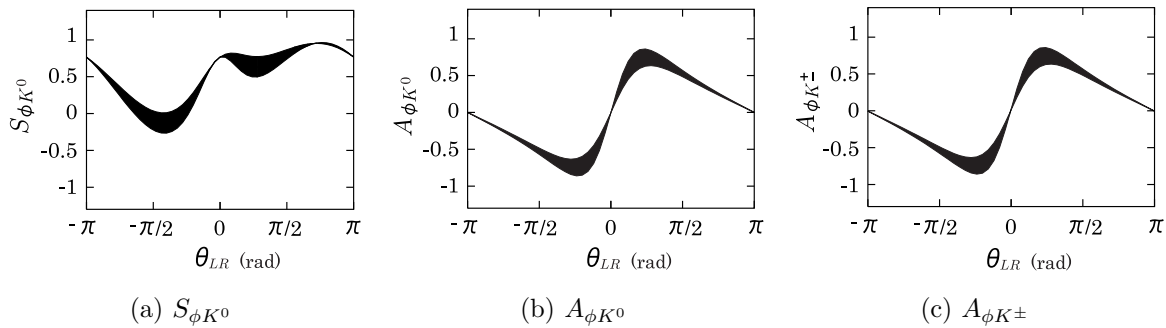}
\caption{%
A possible MSSM modification in $S_{\phi K^0}$, $A_{\phi K^0}$ and
 $A_{\phi K^\pm}$ with the $LR$ mass insertion, which is parameterized as 
 $(\delta^d_{LR})_{23} = -0.015 + r e^{i\,\theta_{LR}}$.  
 Here we take $\phi_1=25^\circ$ and $\phi_3=80^\circ$.  
} 
\label{fig1}
\end{center}
\end{figure}

The numerical results in the case of the $LR$ insertion are displayed in
Fig.~\ref{fig1}. 
Here the $LR$ insertion is parameterized as 
$(\delta^d_{LR})_{23} = -0.015 + r e^{i\,\theta_{LR}}$, and we scan all
values on the allowed region of $r$.  As it can be seen from
Fig.~\ref{fig1}(a), $S_{\phi K^0}$ may deviate significantly from the SM
expectation.  This result is almost the same as that using the QCDF
method~\cite{Kane:2003zi}.  
The result of $A_{\phi K^0}$ and $A_{\phi K^\pm}$ is shown in
Figs.~\ref{fig1}(b) and \ref{fig1}(c), respectively.  $A_{\phi K}$'s
arise from the interference between the penguin amplitudes in the SM and
the CMP ones in the MSSM.  In PQCD, there is a large relative strong
phase between them.  For this reason, $A_{\phi K}$'s may be large
depending on the new physics phase $\theta_{LR}$. It must be noted that
the direct CP asymmetry of the neutral mode has the same tendency as
that of the charged mode, because the CMP contributions as well as the
SM ones are almost the same in both modes.  The current experimental
data of the neutral mode are shown in the introduction, and those of the
charged mode are $A_{\phi K^\pm} = 0.054\pm 0.056\pm 0.012$
(BaBar~\cite{Aubert:2004dy}), $0.01\pm 0.12\pm 0.05$
(Belle~\cite{unknown:2003jf}).  
If we take the result of $A_{\phi K^\pm}$ seriously, there remains only
small room for the allowed region of $\theta_{LR}$ so that the deviation
of $S_{\phi K^0}$ becomes smaller.

\begin{figure}[t]
\begin{center}
\includegraphics*{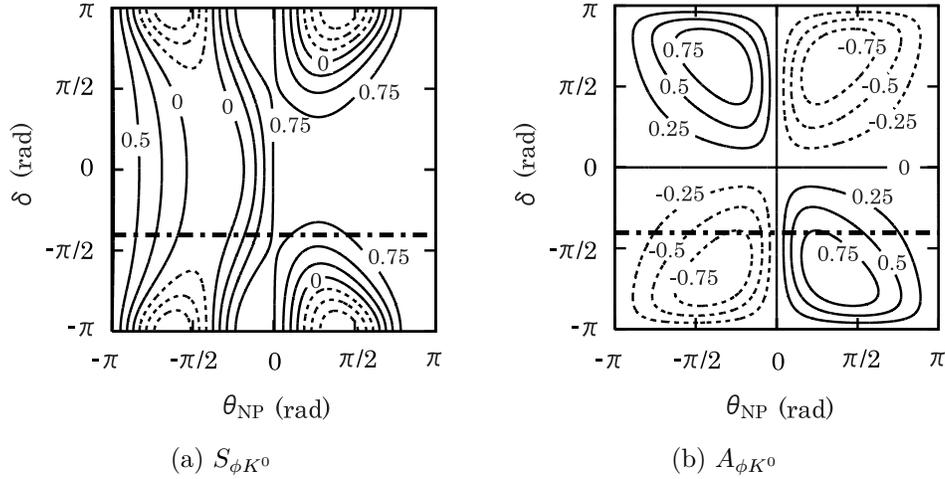}
\caption{%
Contour plots of the CP asymmetries in terms of a new CP violating phase
 $\theta_{\rm NP}$ and a strong phase $\delta$, which is a relative
 phase between the SM and new physics amplitudes.  The magnitude of the
 new physics amplitude is taken to be the central value in the $LR$
 case.  A step between contour lines is $0.25$.  Dotted lines represent
 constant contours with each negative value.  Dot-dashed lines denote the
 PQCD prediction.  
}
\label{fig2}
\end{center}
\end{figure}

Figure~\ref{fig2} shows the strong-phase dependence of the CP
asymmetries.  
Here we parameterize the decay amplitude in terms of a new CP violating
phase $\theta_{\rm NP}$, which is correspond to $\theta_{LR}$, and a
strong phase $\delta$, which is a relative phase between the SM and new
physics amplitudes.  The magnitude of the new physics amplitude is taken
to be the central value in the $LR$ case.  $S_{\phi K^0}$ remains almost
stable in the range of $|\delta|<\pi/2$ as shown in Fig.~\ref{fig2}(a).
As a result, $S_{\phi K^0}$ in our result has the same tendency as that
in the QCDF method. In contrast with $S_{\phi K^0}$, $A_{\phi K^0}$ is
sensitive to the strong phase and the sign of $A_{\phi K^0}$ is flipped
by changing $\delta\to -\delta$ as shown in Fig.~\ref{fig2}(b).  
In consequence, the QCDF prediction has opposite sign from our
result~\cite{Kane:2003zi}.  This fact originates from the difference of
the source of the strong phase between the PQCD and QCDF methods.
Hence we conclude that more theoretical study is needed for the
calculation of the strong phase in order to search for new physics in
the direct CP asymmetry of the $B\to \phi K$ modes.

\section{Acknowledgements}
This work was supported in part by the Research Fellowships of the Japan
Society for the Promotion of Science for Young Scientists, No.13-01722,
and by the Grants-in-aid from the Ministry of Education, Culture,
Sports, Science and Technology, Japan, No.14046201.

\bibliographystyle{plain}

\begin{thebibliography}{99}

\bibitem{Mishima:2003ta}
S.~Mishima and A.~I.~Sanda,
{\it Phys.\ Rev.\ }{\bf D69}, 054005 (2004).

\bibitem{Aubert:2004ii}
B.~Aubert {\it et al.}  [BABAR Collaboration],
{\it Phys.\ Rev.\ Lett.\ }{\bf 93}, 071801 (2004).

\bibitem{Abe:2003yt}
K.~Abe {\it et al.}  [Belle Collaboration],
{\it Phys.\ Rev.\ Lett.\ }{\bf 91}, 261602 (2003).

\bibitem{Aubert:2004dy}
B.~Aubert {\it et al.}  [BABAR Collaboration],
hep-ex/0408072.

\bibitem{Sakai:ichep2004}
Y.~Sakai,
talk given at 
the 32nd International Conference on High Energy Physics
(ICHEP04), Beijing, China, August 2004.

\bibitem{Keum:2000ph}
Y.~Y.~Keum, H.~n.~Li and A.~I.~Sanda,
{\it Phys.\ Lett.\ }{\bf B504}, 6 (2001);
%
{\it Phys.\ Rev.\ }{\bf D63}, 054008 (2001).

\bibitem{Beneke:1999br}
M.~Beneke, G.~Buchalla, M.~Neubert and C.~T.~Sachrajda,
{\it Phys.\ Rev.\ Lett.\ }{\bf 83}, 1914 (1999);
%
        {\it Nucl.\ Phys.\ }{\bf B591}, 313 (2000).

\bibitem{Nagashima:2002ia}
M.~Nagashima and H.~n.~Li,
{\it Phys.\ Rev.\ }{\bf D67}, 034001 (2003).

\bibitem{Chen:2001pr}
C.~H.~Chen, Y.~Y.~Keum and H.~n.~Li,
{\it Phys.\ Rev.\ }{\bf D64}, 112002 (2001);
%
S.~Mishima,
{\it Phys.\ Lett.\ }{\bf B521}, 252 (2001).

\bibitem{Mishima:2003wm}
S.~Mishima and A.~I.~Sanda,
{\it Prog.\ Theor.\ Phys.\ }{\bf 110}, 549 (2003).

\bibitem{Kane:2003zi}
G.~L.~Kane, P.~Ko, H.~b.~Wang, C.~Kolda, J.~h.~Park and L.~T.~Wang,
{\it Phys.\ Rev.\ Lett.\ }{\bf 90}, 141803 (2003).

\bibitem{unknown:2003jf}
K.~-F.~Chen {\it et al.}  [Belle Collaboration],
{\it Phys.\ Rev.\ Lett.\ }{\bf 91}, 201801 (2003).

\end{thebibliography}

\end{document}